\documentclass{article}

\usepackage{arxiv}
\usepackage{amsmath} 
\usepackage[utf8]{inputenc} 
\usepackage[T1]{fontenc}    
\usepackage{hyperref}       
\usepackage{url}            
\usepackage{booktabs}       
\usepackage{amsfonts}       
\usepackage{nicefrac}       
\usepackage{microtype}      
\usepackage{lipsum}
\usepackage{graphicx}
\graphicspath{ {./images/} }

\usepackage{listings}       

\usepackage{enumitem}
\usepackage{algorithm}      
\usepackage{algpseudocode}  
\usepackage{array}     
\newcommand{\method}{EviGraph}  
\newcommand{\tags}{\mathcal{T}} 
\newcommand{\eg}{\mathcal{G}}   
\newcommand{\docs}{\mathcal{D}} 

\title{EviGraph: Towards Verifiable Evidence Construction for Information-Seeking Agents}

\author{
Jiashun Chen, Yirong Mao, Wenhui Que$^{*}$\\
WeChat, Tencent Inc., Beijing, China\\
\{exiajschen, erongmao, victorque\}@tencent.com
}

\begin{document}
\maketitle
\begin{abstract}
Agentic Web search can retrieve relevant information without establishing that the retrieved content actually supports the claims used in an answer. Existing agents typically keep search and evidence recording in a linear interaction trace and optimize primarily for final-answer correctness, providing limited supervision for intermediate grounding. We present \method, a deep-search framework that separates search execution from evidence recording while using a shared policy for the trainable roles. An executor plans concise queries, a frozen evidence verifier inspects source pages and returns verbatim evidence items with an explicit \texttt{polarity}, and the policy maps those items to \texttt{add}/\texttt{support} graph requests that are checked by a deterministic structural validator. The resulting graph serves both as persistent working memory and as a source of dense process rewards, enabling reinforcement learning to directly supervise evidence construction rather than only the final answer. On BrowseComp-Plus, a Qwen3-8B \method{} agent achieves 35.9\% accuracy under a matched interaction budget, compared with 26.9\% for the same dual-role architecture without reinforcement learning and 2.7\% for a monolithic agent, while generating fewer tokens per rollout. Consistent gains on BrowseComp, GAIA, and XBench indicate that explicitly structuring and rewarding evidence recording improves agentic search.
\end{abstract}


\begin{figure}

 \includegraphics[width=\textwidth]{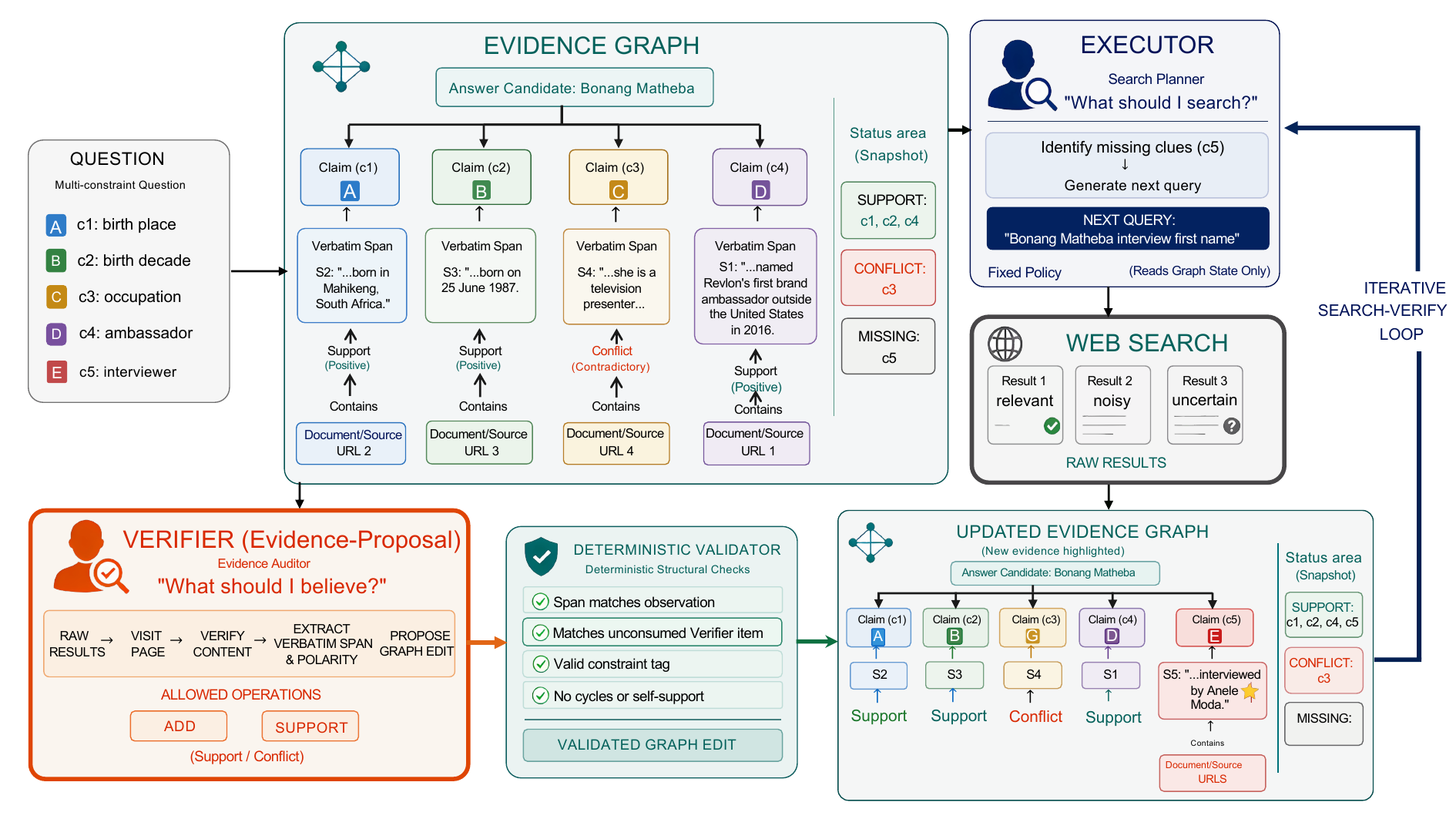}
  \caption{Inference-time \method{} loop. The executor queries from the graph snapshot, the frozen verifier extracts polarity-labeled spans, the evidence-proposal policy requests graph edits, and the deterministic validator commits structurally valid updates.}
  \label{fig:teaser}
\end{figure}

\section{Introduction}
Large language models (LLMs) equipped with search tools can iteratively retrieve external information and solve knowledge-intensive tasks beyond their parametric memory. Recent agentic-search systems train models to decide when to search, formulate queries, consume retrieved passages, and produce final answers using outcome rewards or step-wise retrieval feedback~\cite{chen2025researchlearningreasonsearch,  song-etal-2025-smart}. These methods have substantially improved answer accuracy, but they primarily optimize whether the final answer is correct, while the evidential basis of that answer remains implicit in an ever-growing interaction trajectory.

However, answer correctness does not necessarily imply evidential sufficiency~\cite{min-etal-2023-factscore, gao-etal-2023-rarr}. First, retrieval relevance is not evidential support: a page may mention the correct entity while referring to a different time period, setting, geographic scope, or numerical value. Second, answer-level rewards can encourage shortcut behavior, allowing a correct answer despite missing, contradictory, or unsupported evidence. Third, citations alone do not reveal which source span supports a particular claim~\cite{gao-etal-2023-enabling}. These limitations point to a missing component in current search agents: an explicit representation of \emph{which evidence supports which claim, under which constraint, and how multiple pieces of evidence jointly establish the final conclusion}.

This issue is pronounced in Web research tasks that require satisfying multiple independently searchable constraints~\cite{chen-etal-2026-browsecomp}. A candidate page may satisfy an entity-type constraint while violating a temporal or geographic one, whereas another page satisfies a different subset of requirements. Retrieving relevant pages therefore does not establish that a candidate satisfies all constraints; solving such tasks requires an agent to \emph{construct evidence that jointly satisfies the required constraints}.

Based on this observation, we introduce \textbf{EviGraph}, which reformulates agentic search as \emph{evidence construction}: a heterogeneous DAG anchors atomic claims and answer candidates to source documents through verbatim spans and polarity-labeled evidence edges. Unlike prior evidence graphs used mainly as passive state or external supervision~\cite{zhou-etal-2019-gear}, EviGraph uses this same span-grounded structure as both working memory and a source of process rewards.

We decouple deep search into two trainable roles under a shared policy: an \emph{executor} issues concise queries, while an \emph{evidence-proposal role} decides how verified evidence should update the graph. A frozen evidence verifier reads each raw result or visited page and returns candidate evidence items containing a source, verbatim span, constraint tag, relevance explanation, and \texttt{polarity}; it never edits the graph. The evidence-proposal role maps each item to an \texttt{add} request for a new candidate or a \texttt{support} request for an existing one, where the edge polarity determines whether the item supports or conflicts with the claim. A deterministic structural validator checks the request and graph invariants but makes no semantic entailment or answer-stopping decision. Both roles are realized by the same policy model $\pi_\theta$ and distinguished only by role context, so search and evidence-recording proposals are jointly learned without treating the frozen verifier as a trainable policy.

Span anchoring and explicit constraint tags make graph quality measurable at each step, yielding dense rewards for valid evidence construction rather than only final-answer correctness. Agentic search thus becomes a structured sequential decision process that progressively constructs and composes auditable evidence.

Our contributions are as follows:
\begin{itemize}[leftmargin=*,nosep]
    \item \textbf{Agentic search as evidence construction.}
    We formulate agentic search as constructing a typed, span-grounded claim--evidence graph whose validated structure supplies dense process rewards.

    \item \textbf{Shared-policy search and evidence recording.}
    The executor and the evidence-proposal role are two contexts of a single policy model $\pi_\theta$. The learned policy proposes searches and graph-writing requests, whereas a frozen verifier supplies polarity-labeled evidence items and a deterministic validator checks only syntax and graph invariants, making explicit which parts are learned and which are fixed at execution time.

    \item \textbf{Role-specific reinforcement learning with reward-hacking safeguards.}
    We provide an RL formulation with separate process rewards for executor search decisions and policy-generated evidence-recording proposals, plus a shared final-answer reward. Coverage uses only positive-polarity evidence, while unresolved support--conflict pairs are penalized. Structural validation, a potential-difference reward in the policy-invariant form $\gamma\Phi(\eg')-\Phi(\eg)$~\cite{ng1999policy}, canonical evidence signatures, a per-candidate--tag--polarity acceptance-bonus cap, and a KL penalty jointly mitigate reward farming from duplicate or superficially varied records.

    \item \textbf{Evaluation on text-only and multimodal Web search.}
    Controlled Qwen3-8B experiments on BrowseComp-Plus~\cite{chen-etal-2026-browsecomp} isolate the contributions of role decomposition and reinforcement learning, and the gains carry over to BrowseComp, GAIA, and XBench. We additionally show that the same inference-time loop transfers to multimodal Web search on LiveVQA~\cite{fu2025seekingupdatinglivevisual}, using a pretrained Qwen3-VL-8B~\cite{bai2025qwen3vltechnicalreport} policy without additional fine-tuning.
\end{itemize}

\section{Related Work}

\paragraph{Agentic search and retrieval learning.}
Search-R1~\cite{searchr1}, R1-Searcher~\cite{song-etal-2025-smart}, StepSearch~\cite{stepsearch}, ZeroSearch~\cite{zerosearch}, and DeepResearcher~\cite{deepresearcher} train language models to interleave reasoning with retrieval. Their rewards primarily measure tool use, information gain, or final-answer quality. \method{} instead treats the evidence structure itself as the process-supervision signal, so a retrieved passage is useful only when it grounds a specific claim.

\paragraph{Evidence graphs and verifiable reasoning.}
 MedCEG~\cite{medceg} derives critical-evidence graphs from annotations. AutoGraph-R1~\cite{autographr1} and GraphPRM~\cite{graphprm} optimize graph construction or graph-reasoning rewards. In contrast, \method{} induces a span-anchored graph online from noisy Web pages and uses the same validated graph both as working memory and as a self-supervised reward source.

\paragraph{Deep-research and memory agents.}
Multimodal research agents such as OpenSearch-VL~\cite{opensearchvl} combine search, browsing, and OCR, while memory systems such as Mem0~\cite{mem0}, MemGPT~\cite{memgpt}, A-MEM~\cite{amem}, and REAL~\cite{real} compress long histories for later retrieval. These systems improve access to information but generally retain passages or summaries as opaque context. \method{} is complementary: it converts retrieved content into typed, auditable claim--evidence relations before answering.

\section{Method}
\label{sec:method}

\subsection{Problem Formulation}
Given a question $q$, the system first identifies a set of required constraints $\tags=\{k_1,\ldots,k_m\}$. At step $t$, it has observed documents $\docs_t$ and maintains an evidence graph $\eg_t=(V_t,E_t)$. The objective is to produce an answer $a$ while constructing a graph in which each recorded claim is traceable to a verbatim source span and each required constraint has positive support, conflicting evidence, or remains unresolved. The graph itself does not assign solved, contradicted, or closed statuses.

\subsection{Evidence-Graph Schema}
\label{sec:structure}
Table~\ref{tab:schema} summarizes the graph schema. The node set partitions into documents $V_D$, verbatim spans $V_S$, candidate--constraint claims $V_A$, and answer candidates $V_F$. A \textsc{Contains} edge links a document to a span, while an evidence edge links a span to a claim with $\texttt{polarity}\in\{\texttt{support},\texttt{conflict}\}$; scope and source context remain fields of the evidence record. Answer candidates group the atomic claims considered by the executor, which decides sufficiency from their positive support, unresolved constraints, and recorded conflicts rather than from graph-level solved or refuted statuses.

The schema is modality-agnostic: visual evidence is converted to text before entering the graph. On multimodal benchmarks, image regions, OCR output, and crop boxes are rendered as textual descriptions, and each description is treated as an ordinary span anchored to its source page, using the same \textsc{Contains} and evidence edges as quoted text. No dedicated image nodes or visual-anchoring rules are therefore needed, and the validator, rewards, and provenance invariant are unchanged across modalities.

\begin{table}[t]
\centering
\small
\caption{Node and edge types in the evidence graph.}
\label{tab:schema}
\begin{tabular}{@{}p{0.25\columnwidth}p{0.69\columnwidth}@{}}
\toprule
Type & Semantics \\
\midrule
Document & Opened source with a stable identifier \\
Span & Verbatim text anchored to a document \\
Atomic claim & One candidate--constraint assertion \\
Answer candidate & Candidate identifier referenced by atomic claims \\
\midrule
\textsc{Contains} & A document contains a quoted span ($E_{\mathrm{contains}}\subseteq V_D\times V_S$) \\
Evidence & A span is linked to a claim with $\texttt{polarity}\in\{\texttt{support},\texttt{conflict}\}$ \\
\bottomrule
\end{tabular}
\end{table}

The schema makes partial evidence explicit. Suppose one page establishes that an institution held an event in 2002 and another establishes its location, but no source verifies the required 2003 ceremony. A linear trace may encourage merging these facts into a complete answer; in \method{} the first two facts create support paths for their tags while the ceremony tag remains unresolved, so coverage, contradiction, and provenance can be inspected independently of the final answer. Each accepted claim stores its normalized text, candidate identifier, constraint tag, and incoming polarity-labeled evidence edges; a span stores the source identifier, verbatim quote, and character offsets when available.

\subsection{Sequential Dual-Role Decision Process}
\label{sec:roles}
We formulate construction as a finite-horizon partially observed decision process with shared environment state
\begin{equation}
S_t=(q,\tags,\eg_t,\mathcal{L}_t,\mathcal{O}_t,B_t),
\end{equation}
where $\mathcal{L}_t$ is the query log, $\mathcal{O}_t$ the current retrieval observations, and $B_t$ the remaining tool budget. The executor and the evidence-proposal role are functionally distinct contexts of one policy $\pi_\theta$. The executor observes $(q,\tags,\operatorname{render}(\eg_t),\mathcal{L}_t,B_t)$ and chooses $a_t^E\in\{\textsc{Search}(u),\textsc{Answer}(a)\}$ without inspecting raw pages or modifying the graph. It first locates candidates, then targets unresolved constraints, pivoting after unproductive searches.

The frozen verifier inspects results, opens promising links, and returns a verdict plus zero or more items containing a candidate, tag, source, verbatim span, relevance explanation, and polarity. The evidence-proposal context maps each non-noise item exactly once to $a_t^P\in\{\textsc{Add},\textsc{Support}\}$ without changing its fields: \textsc{Add} introduces a candidate with its first item, and \textsc{Support} attaches an item to an existing candidate. Page inspection and polarity judgment thus remain verifier decisions, graph writing remains a policy decision, and answer stopping remains an executor decision.

A proposal induces $\tilde{\eg}_{t+1}=f(\eg_t,a_t^P)$. The deterministic validator commits it only if it matches an unconsumed verifier item and preserves the graph invariants; otherwise the graph is unchanged and the policy receives a typed error. The updated snapshot lists positive support, conflicts, and a \texttt{MISSING} set computed only from positive-polarity evidence, enabling the executor to target unresolved or disputed constraints without rereading the trajectory.

This separation exposes three failure sources that a monolithic trace conflates: \emph{retrieval failure} when no useful source is found, \emph{recording failure} when useful evidence yields no valid request, and \emph{decision failure} when the executor stops too early or searches after sufficient evidence is available. Each corresponds to a distinct transition and receives role-specific credit.

\begin{algorithm}[t]
\caption{Dual-role evidence construction.}
\label{alg:search}
\begin{algorithmic}[1]
\Require question $q$, required tags $\tags$, budget $B$
\State $\eg\leftarrow\emptyset$; query log $\mathcal{L}\leftarrow\emptyset$
\While{$B$ is not exhausted}
  \State executor chooses \textsc{Search}$(u)$ or \textsc{Answer}$(\eg)$
  \If{\textsc{Answer} is chosen} \State \Return answer \EndIf
  \State results $\leftarrow \textsc{Search}(u)$; $\mathcal{L}\leftarrow\mathcal{L}\cup\{u\}$
  \State frozen verifier returns polarity-labeled evidence items or a concrete noise summary
  \State evidence-proposal policy maps items to \texttt{add} or \texttt{support}, if any
  \State structural validator accepts valid requests and returns the updated snapshot
\EndWhile
\State \Return answer from the current graph
\end{algorithmic}
\end{algorithm}

\subsection{Structural Validation and Termination}
\label{sec:converge}
Before a graph-writing request is committed, the deterministic validator checks that (i) the cited span occurs in the referenced observation, (ii) the source and node identifiers exist, (iii) the constraint label belongs to $\tags$, (iv) the tag, source, span, and polarity exactly match an unconsumed verifier item, (v) the request introduces neither a cycle nor self-support, and (vi) an \texttt{add} or \texttt{support} request is well-formed. It also computes a canonical evidence signature from the source identifier, span offsets, normalized claim, tag, and polarity, converting exact duplicates into zero-reward no-ops. Dropping a non-noise verifier item or changing its polarity is treated as a rejected proposal, and rejected requests are returned with a structured error code.

The validator enforces graph integrity but decides neither semantic entailment nor termination. The executor may answer when its answer prompt judges that the recorded evidence pins down a single candidate; it may also seek corroboration or keep searching. The graph therefore retains all recorded support and conflicting quotes, and uncertainty is exposed through missing constraints rather than hidden behind a validator-issued verdict.

\paragraph{Source-reachability invariant.}
Starting from an empty graph, every accepted claim is reachable from an observed source span by a directed evidence path, irrespective of polarity. The property holds inductively: \texttt{add} requires an existing source and verbatim span, so the new claim is immediately reachable from that source; \texttt{support} attaches another observed span to an existing claim, preserving reachability. Since the validator also rejects missing identifiers, invalid tags or polarity, duplicate signatures, cycles, and self-support, structural hallucinations such as invented URLs or circular self-support cannot enter the persistent graph. This does not prove that a source is true or that a span semantically entails a claim, but it guarantees that every recorded belief is auditable against an actual observation.

\subsection{Handling Noisy Retrieval}
A visit-first policy prevents the frozen verifier from treating incomplete snippets as evidence: relevant but insufficient results are opened and inspected, irrelevant or empty results are summarized as noise, and promising links are returned as leads. Temporal, geographic, and numerical scope is preserved in the constraint tag and \texttt{why\_relevant} fields. Contradictory spans remain in the graph as polarity-\texttt{conflict} edges, are shown separately in the snapshot, and do not count toward coverage; no \texttt{modify}, \texttt{delete}, or \texttt{resolve} operation is used.

\subsection{Worked Example}
\label{subsec:illustrative}
We show the complete interface transition for the first round and abbreviate the rest; long spans are shortened for display, while stored records retain the verbatim text.
\begin{lstlisting}
Question: identify the interviewer's first name for a woman satisfying
c1=birth_place, c2=birth_decade, c3=occupation,
c4=revlon_ambassador, and c5=interviewer_first_name.
Initial: MISSING={c1,c2,c3,c4,c5}.

Round 1 (full transition)
Executor query: "Revlon first brand ambassador outside United States"
Verifier item:
  candidate="Bonang Matheba", tag=c4, polarity=support,
  S1="... named Revlon's first brand ambassador outside the United
  States in 2016."
Policy request: add(C,S1,c4,support); Validator: accepted
Graph: C=Bonang Matheba; MISSING={c1,c2,c3,c5}.

Remaining rounds (abridged)
R2: S2 "born in Mahikeng, South Africa, on 25 June 1987" covers c1,c2.
R3: S3 lists television, radio, acting, business; support(C,S3,c3).
R4: S4 records "Adaku" as the interviewer; support(C,S4,c5).
MISSING={}; Executor: <answer>Adaku</answer>
\end{lstlisting}

\section{Learning}
\label{sec:training}

\subsection{Macro-Step Credit Assignment}
\label{sec:reward}
A search round is treated as one macro-step. Starting from graph $\eg_m$, the executor emits a query or a final answer. For a query, the environment retrieves results, the frozen verifier performs zero or more visits, and the policy emits zero or more \texttt{add}/\texttt{support} requests, producing $\eg_{m+1}$. The executor is therefore scored on the complete transition $\eg_m\rightarrow\eg_{m+1}$ rather than on the query string alone.

For candidate $f\in V_F$, let $\mathcal{T}^{\mathrm{cov}}(\eg,f)\subseteq\tags$ be the tags whose candidate--constraint claim has at least one validator-accepted \texttt{support} edge; conflict edges are excluded. Define candidate coverage by
\begin{equation}
c(\eg)=
\max_{f\in V_F}
\frac{|\mathcal{T}^{\mathrm{cov}}(\eg,f)|}{\max(1,|\tags|)},
\label{eq:cov}
\end{equation}
with $c(\eg)=0$ when $V_F=\emptyset$. Taking the maximum per candidate prevents evidence for different candidates from being merged into artificial coverage. Let $\mathcal{T}^{\mathrm{ret}}_m$ be the required tags for which the frozen verifier reports relevant evidence in round $m$, i.e.\ the tags appearing in its emitted items, so novelty depends on the verifier's relevance decisions rather than on the executor's own claim about its query. Retrieval novelty is
\begin{equation}
n_m=
\frac{|\,\mathcal{T}^{\mathrm{ret}}_m
\setminus\bigcup_{\ell<m}\mathcal{T}^{\mathrm{ret}}_\ell\,|}
{\max(1,|\tags|)}.
\end{equation}
For a search action, the executor process reward is
\begin{equation}
r_m^E=
\alpha n_m+
\bigl(\gamma\beta\,c(\eg_{m+1})-\beta\,c(\eg_m)\bigr)
-\kappa_s .
\label{eq:rewe}
\end{equation}
An answer action receives no search cost; its quality is handled by the terminal reward. The coverage term is potential-based shaping of the bounded potential $\Psi(\eg)=\beta\,c(\eg)$ in the discount-consistent form $\gamma\Psi(\eg_{m+1})-\Psi(\eg_m)$~\cite{ng1999policy}, assigning the delayed coverage gain to the query that produced the evidence without adding a separately farmable signal.

\subsection{Incremental Evidence-Proposal Reward}
\label{sec:proprew}
For a valid graph $\eg$ over node sets $V_D,V_S,V_A,V_F$, let $V_A^{\pm}$ denote claims with both support and conflict evidence, and let $V_A^e$ denote claims with any incoming evidence. The unresolved-conflict rate is
\begin{equation}
U(\eg)=\frac{|V_A^{\pm}|}{\max(1,|V_A^e|)},
\label{eq:U}
\end{equation}
and, letting $V_{\mathrm{off}}$ contain the claim and candidate nodes ($V_A\cup V_F$) whose constraint tag is not in $\tags$, the off-task penalty
\begin{equation}
O(\eg)=\frac{|V_{\mathrm{off}}|}{\max(1,|V_A\cup V_F|)}.
\label{eq:O}
\end{equation}
Restricting both numerator and denominator to claim and candidate nodes keeps the ratio well defined: documents and spans are never linked to a tag directly, so a newly retrieved span that has not yet been attached to a claim is never counted as off-task and exploratory retrieval is not penalized.
Polarity is supplied as a dedicated verifier-output field and cannot be inferred solely from free-form \texttt{why\_relevant} text. Both $U$ and $O$ are $0$ when their denominator is $0$, and an atomic claim with several conflict records contributes once. Coverage $c(\eg)$ is defined in Eq.~\eqref{eq:cov}; all three terms lie in $[0,1]$. Define the bounded graph-quality potential
\begin{equation}
\Phi(\eg)=
\lambda_{\mathrm{cov}}c(\eg)-
\lambda_u U(\eg)-
\lambda_o O(\eg),
\label{eq:Phi}
\end{equation}
with terminal potential $\Phi(\eg_{\mathrm{end}})=0$.

\paragraph{What the conflict term penalizes.}
$U(\eg)$ penalizes claims left in an \emph{unresolved} support--conflict state, not the act of recording a conflict. The intended gradient is to keep searching until a disputed constraint is corroborated by an additional source, which raises $c(\eg)$, rather than to suppress the conflicting quotation. Suppression is also not the cheaper option: dropping a non-noise verifier item counts as a rejected proposal and incurs $-\lambda_v$ (Eq.~\eqref{eq:rewv}) while leaving $\Phi$ unchanged, so the return cannot be improved by hiding contradictory evidence. Since a claim with several conflict records contributes to $|V_A^{\pm}|$ only once, further sources for an already-disputed claim do not compound the penalty. We keep $\lambda_u<\lambda_{\mathrm{cov}}$ (Table~\ref{tab:hyper}) so that resolving a constraint is always worth more than avoiding a dispute.

For evidence-proposal substep $j$ in round $m$, let $\eg_{m,j}$ and $\eg_{m,j+1}$ be the graphs before and after the proposed request. The frozen verifier supplies the immutable evidence item, the shared policy maps it to a request, and the validator decides only whether the request satisfies the structural invariants. Its reward is
\begin{equation}
r_{m,j}^P=
\lambda_v\,v_{m,j}
+\bigl(\gamma\Phi(\eg_{m,j+1})-\Phi(\eg_{m,j})\bigr)
\label{eq:rewv}
\end{equation}
where $v_{m,j}=1$ only when an accepted request records the first item for a candidate--tag--polarity triple, $v_{m,j}=-1$ for a structurally rejected request, and $v_{m,j}=0$ for noise, an exact duplicate, or additional corroboration of a triple that already received its bonus. The acceptance bonus is thus capped once per triple, and a rejected or duplicate request leaves the graph unchanged and yields zero potential change. Canonical signatures, the per-triple bonus cap, the finite graph, and the search budget jointly \emph{mitigate} reward farming from duplicate or superficially varied records; we do not claim they eliminate every possible exploit.

At episode end, if a gold answer $y$ is available, we use
\begin{equation}
R_{\mathrm{final}}=
\mathbb{1}[\operatorname{norm}(\hat a)=\operatorname{norm}(y)].
\end{equation}
The equivalence test is computed during training by a frozen DeepSeek-V4-Flash~\cite{deepseekv4} scorer, which receives only the predicted and gold answers and returns a binary label. This training-time scorer is distinct from the evidence verifier: it produces no evidence items, never inspects pages or edits the graph, contributes nothing to the process rewards of Eqs.~\eqref{eq:rewe} and~\eqref{eq:rewv}, and is unused at inference time and in all reported evaluations. For instances without a gold answer, we set the terminal-reward weight to zero rather than constructing a pseudo-label.

Let $r_\ell^r$ be the ordered process rewards received by role $r\in\{E,P\}$ and let $L_r$ be the number of actions taken by that role. The return for its $\ell$-th action is
\begin{equation}
G_\ell^r=
\sum_{k=\ell}^{L_r-1}\gamma^{k-\ell}r_k^r
+\gamma^{L_r-\ell}\lambda_fR_{\mathrm{final}}.
\label{eq:reward}
\end{equation}
The terminal reward is thus added once through the return, avoiding a bias toward longer episodes.

\paragraph{Why the shaping term is not degenerate.} With $\gamma=1$ the potential differences in Eqs.~\eqref{eq:rewe} and~\eqref{eq:rewv} telescope along a fixed rollout, so a full-episode sum would collapse to $-\Phi(\eg_0)$. Two choices prevent this degeneracy. First, the reset $\Phi(\eg_{\mathrm{end}})=0$ is an accounting convention for the final transition and is not emitted as an extra reward, so no spurious episode-level bonus appears. Second, GRPO compares candidates branching from the same state and role, each completed by an independent $\pi_{\theta_{\mathrm{old}}}$ rollout; candidates therefore reach different terminal graphs and lengths, so the accumulated shaping term differs across the group and remains informative for the advantage in Eq.~\eqref{eq:adv}. Shaping redistributes credit within an episode rather than adding a farmable quantity~\cite{ng1999policy}.

\subsection{Role-Conditioned GRPO}
\label{sec:objective}
The executor and evidence-proposal policy share parameters but use different role tokens and action spaces. For one state $s$ assigned to role $r$, we sample $N_r$ candidate action sequences $\{a_i\}_{i=1}^{N_r}$ and evaluate each in the environment: executor candidates run one complete search--evidence-recording macro-step, while evidence-proposal candidates are applied to independent graph copies and checked by the validator. The remaining episode is then rolled out with $\pi_{\theta_{\mathrm{old}}}$ so that the return $G_i^r$ includes delayed process and terminal rewards. Advantages are normalized only within candidates from the same state and role:
\begin{equation}
\hat A_i^r=
\frac{G_i^r-\mu_{s,r}}{\sigma_{s,r}+\epsilon},
\quad
\mu_{s,r}=\frac{1}{N_r}\sum_iG_i^r.
\label{eq:adv}
\end{equation}
Here $\sigma_{s,r}$ is the standard deviation of the same state--role group. For token $k$ of candidate $a_i$, define
\begin{equation}
\rho_{i,k}^r(\theta)=
\frac{\pi_\theta(a_{i,k}\mid s,a_{i,<k},r)}
{\pi_{\theta_{\mathrm{old}}}(a_{i,k}\mid s,a_{i,<k},r)}.
\end{equation}
The clipped loss for role $r$ is
\begin{equation}
\begin{split}
\mathcal{L}_r=-\mathbb{E}\Bigg[
\frac{1}{N_r}\sum_{i=1}^{N_r}\frac{1}{|a_i|}
\sum_{k=1}^{|a_i|}
\min\Big(&\rho_{i,k}^r\hat A_i^r,\\
&\operatorname{clip}(\rho_{i,k}^r,1-\epsilon,1+\epsilon)
\hat A_i^r\Big)\Bigg].
\end{split}
\end{equation}
The shared-policy objective is
\begin{equation}
\mathcal{L}=
\omega_E\mathcal{L}_E+
\omega_P\mathcal{L}_P+
\beta_{\mathrm{KL}}\,
\mathbb{E}_{s,r}
\!\left[
D_{\mathrm{KL}}\!\left(
\pi_\theta(\cdot\mid s,r)
\Vert
\pi_{\mathrm{ref}}(\cdot\mid s,r)
\right)\right].
\label{eq:grpo}
\end{equation}
$\omega_E$ and $\omega_P$ control the contribution of the two roles; they are kept fixed across the compared training runs.
\section{Experiments}

We evaluate \method{} on five benchmarks covering diverse web-search and multimodal search scenarios. BrowseComp-Plus~\cite{chen-etal-2026-browsecomp}, BrowseComp~\cite{wei2025browsecompsimplechallengingbenchmark}, GAIA~\cite{mialon2023gaiabenchmarkgeneralai}, and XBench~\cite{chen2025xbenchtrackingagentsproductivity} evaluate general web search, while LiveVQA~\cite{fu2025seekingupdatinglivevisual} evaluates multimodal web search with visual inputs. Together they cover both fixed-corpus and live-web settings.

\paragraph{Models and Search Environment.}
We conduct experiments with two model sizes. Qwen3-8B-Base~\cite{yang2025qwen3technicalreport} serves as the backbone for our main experiments, including both EviGraph construction and GRPO~\cite{shao2024deepseekmathpushinglimitsmathematical} training. We additionally evaluate \method{} with the larger Qwen3-32B-Base~\cite{yang2025qwen3technicalreport} to assess the scalability of the proposed framework. For multimodal LiveVQA, we use the pretrained Qwen3-VL-8B~\cite{bai2025qwen3vltechnicalreport} as the policy model without any additional fine-tuning: it is applied in a zero-shot setting and only its prompt and search environment differ from the baselines. For BrowseComp-Plus~\cite{chen-etal-2026-browsecomp}, we use Qwen3-Embedding-8B~\cite{zhang2025qwen3embeddingadvancingtext} to construct the retrieval index over the benchmark's fixed document corpus. For live-web search, we use the Serper API to provide access to external web search.

\paragraph{Evaluation Protocol.}
For BrowseComp-Plus~\cite{chen-etal-2026-browsecomp}, we follow the official evaluation protocol over its 830-question evaluation set and use Qwen3-32B~\cite{yang2025qwen3technicalreport} as the LLM judge to assess whether the generated answer is consistent with the reference answer. We report answer accuracy as the primary metric.

For BrowseComp~\cite{wei2025browsecompsimplechallengingbenchmark}, GAIA~\cite{mialon2023gaiabenchmarkgeneralai}, and XBench~\cite{chen2025xbenchtrackingagentsproductivity}, we follow the open-web evaluation protocol of OpenResearcher~\cite{li2026openresearcherfullyopenpipeline}. These benchmarks are evaluated in a live-web setting, where the agent iteratively interacts with the Serper API to issue web searches, reads the retrieved information, and produces a final answer. Following OpenResearcher, we use GPT-4.1 as an LLM judge to compare each prediction with the provided reference answer and report judge-based answer accuracy on each benchmark.

For LiveVQA, we follow the evaluation protocol of Vision-DeepResearch~\cite{fu2025seekingupdatinglivevisual}, using the identical evaluation set, the same tool environment (text search, image search, image cropping, and web browsing), and the same judge, Qwen3-VL-30B-A3B-Instruct, so that the comparison with reported baselines is sound.

\subsection{Training Data and Procedure}
The training prompts are derived from the Search-R1 question--query data~\cite{searchr1}, which supplies 169,615 question--answer pairs from Natural Questions (NQ) and HotpotQA~\cite{hotpotqa} (79,168 and 90,447 examples, respectively). All evaluation configurations share the same agent loop, tool permissions, retrieval backend, and interaction budget; they differ only in the control architecture (single-role vs.\ dual-role) and in whether the policy is updated. A question defines the episode goal, and its associated query is used as an initial retrieval seed. We apply no supervised cold-start or format fine-tuning; during GRPO~\cite{shao2024deepseekmathpushinglimitsmathematical} the executor samples its own queries, and subsequent retrieval, verifier visits, and policy graph-edit requests are generated online. Because Search-R1~\cite{searchr1} supplies no evidence-graph labels, all process rewards are computed from observed pages and validator-accepted graph transitions. The no-RL baselines run the base model directly as an agent without any parameter update, and rejected requests remain visible to the reward rather than being removed from the training distribution.


\begin{figure*}[t]
\centering
\includegraphics[width=\textwidth]{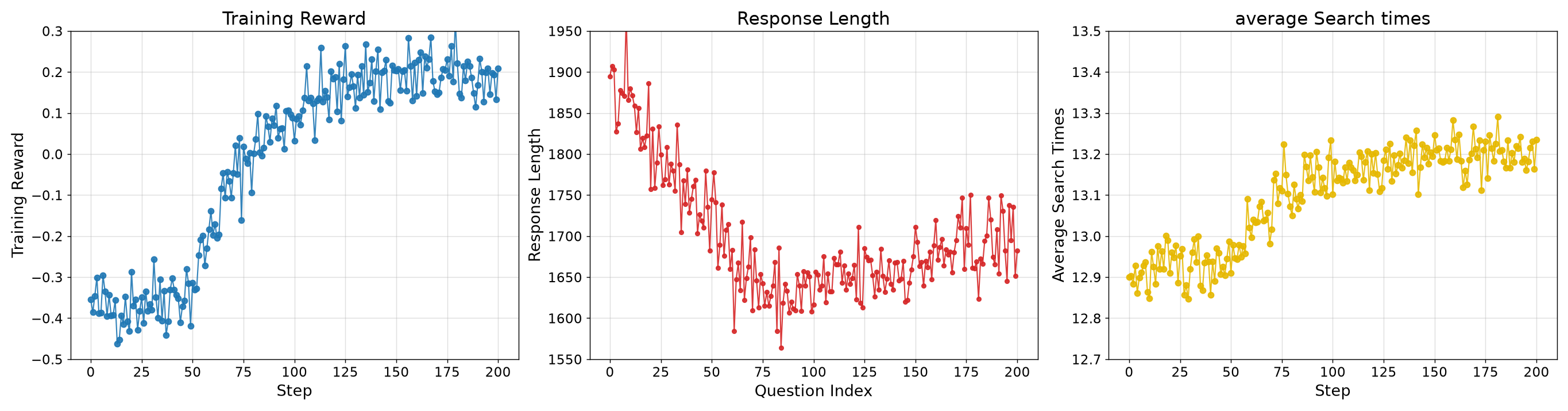}
\caption{Training dynamics: reward (left), generated response length (center), and mean searches per rollout (right).}
\label{fig:train}
\end{figure*}

\subsection{Implementation Details}
\label{sec:impl}
The RL configuration optimizes the role-conditioned GRPO objective (Eq.~\eqref{eq:grpo}), while the no-RL baselines run the same loop without updating the policy. Span grounding, relevance judgments, and evidence extraction are performed by a frozen Qwen3 verifier using a fixed instruction prompt: we use Qwen3-8B or Qwen3-32B according to the model configuration and keep its parameters fixed throughout training and evaluation, while the structural validator checks only the JSON schema and graph invariants. As stated in Section~\ref{sec:reward}, DeepSeek-V4-Flash~\cite{deepseekv4} is used only during GRPO training to score terminal-answer correctness, and plays no role in evidence extraction, process rewards, or inference.
We fine-tune Qwen3-8B-Base with AdamW ($\beta_1{=}0.9$, $\beta_2{=}0.95$, weight decay $0.01$) at a learning rate of $1{\times}10^{-6}$, using a cosine schedule with 10 warmup steps for 1,200 steps. For each prompt and role, we sample $N_r{=}8$ rollouts as the advantage group. Training runs on $8\times$ NVIDIA H20 GPUs (96\,GB HBM each), with one prompt per GPU and gradient accumulation of 4. Responses are truncated at 4,096 tokens, and generation uses temperature 1.0 with top-$p$ 0.95. Each episode is capped at 50 LLM interaction turns across all ablation configurations, so search-count differences reflect the models' stopping behavior rather than unequal maximum budgets.
GRPO training proceeds in two phases: process-only rewards first ($\lambda_f{=}0$), then the terminal reward enabled for instances with gold answers; there is no supervised cold-start stage. Reported BrowseComp-Plus numbers are the mean over three training seeds with standard deviation below 0.5 points; the remaining benchmarks are evaluated once per configuration.

\paragraph{Reward and optimization hyperparameters.}
All coefficients are fixed across the compared training runs; the values are listed in Table~\ref{tab:hyper}. Here $\beta$ is the coverage-potential weight (distinct from the AdamW coefficients $\beta_1,\beta_2$), and $\gamma$ is the reward discount shared by Eqs.~\eqref{eq:rewe}--\eqref{eq:reward}.
\begin{table}[t]
\centering
\caption{Reward coefficients and selected optimization hyperparameters.}
\label{tab:hyper}
\begin{tabular}{lll}
\toprule
Symbol & Meaning & Value \\
\midrule
$\alpha$ & executor retrieval-novelty weight & $0.2$ \\
$\beta$ & executor coverage-potential weight & $0.5$ \\
$\kappa_s$ & executor per-search cost & $0.05$ \\
$\lambda_u$ & unresolved-conflict penalty in $\Phi$ & $0.2$ \\
$\lambda_{\mathrm{cov}}$ & coverage weight in $\Phi$ & $0.5$ \\
$\lambda_o$ & off-task penalty weight in $\Phi$ & $0.2$ \\
$\lambda_v$ & evidence-proposal acceptance bonus & $0.1$ \\
$\lambda_f$ & terminal-reward weight & $1.0$ \\
$\gamma$ & reward discount & $1.0$ \\
$\beta_{\mathrm{KL}}$ & KL penalty weight & $0.04$ \\
\bottomrule
\end{tabular}
\end{table}

\paragraph{Episode accounting.}
The reported token statistic is the mean generated-token count per evaluation rollout, including executor queries, policy graph-edit requests, and frozen-verifier evidence records, but excluding environment-supplied page text. Search count is reported per rollout and records executor calls; page visits are verifier actions and are not conflated with searches. Final answers are scored by the benchmark-specific judges above: Qwen3-32B for BrowseComp-Plus, GPT-4.1 for BrowseComp, GAIA, and XBench, and Qwen3-VL-30B-A3B-Instruct for LiveVQA. The graph remains available for auditing but does not alter answer scoring.

\subsection{Main Results}
Table~\ref{tab:bcplus-results} reports BrowseComp-Plus results. \method{} reaches 35.9\%, outperforming the Qwen3-32B ReAct baseline (10.4) and the Qwen3-8B ReAct baseline (8.2) on the same 830-question evaluation set, while remaining below the strongest closed and specialized systems. The comparison is therefore most informative as a parameter-efficient open-model result rather than a state-of-the-art claim.

\begin{table}[t]
\centering
\caption{Accuracy (\%) on the 830-question BrowseComp-Plus evaluation set; the best result in each block is bold.}
\label{tab:bcplus-results}
\begin{tabular}{lc}
\toprule
Method & Accuracy \\
\midrule
GPT-4.1~\cite{openai2025gpt41} & 36.4 \\
Claude-4-Opus~\cite{anthropic2025claude4} & 36.8 \\
Gemini-2.5-Pro~\cite{comanici2025gemini25pushingfrontier} & 29.5 \\
Kimi-K2~\cite{kimiteam2026kimik2openagentic} & 35.4 \\
DeepSeek-R1~\cite{Guo_2025} & 16.4 \\
Nemotron-3-Nano~\cite{nvidia2025nemotron3nanoopen} & 20.8 \\
Qwen3-8B~\cite{yang2025qwen3technicalreport} + ReAct & 8.2 \\
Qwen3-32B~\cite{yang2025qwen3technicalreport} + ReAct & 10.4 \\
Tongyi DeepResearch~\cite{deepresearcher} & \textbf{44.5} \\
CutBill-30B-A3B~\cite{wu2025cut} & 30.3 \\
\midrule
\method{} (Qwen3-8B) & 26.9 \\
\method{} (Qwen3-8B RL) & 35.9 \\
\method{} (Qwen3-32B) & \textbf{36.6} \\
\bottomrule
\end{tabular}
\end{table}

\subsection{LiveVQA Results}
Table~\ref{tab:livevqa} reports the multimodal comparison. The full Qwen3-VL-8B \method{} system reaches 78.0\%, improving by 27.0 points over the no-search variant, in which the executor performs no retrieval and builds the graph only from evidence already present in the prompt, and by 31.0 points over direct inference. It exceeds Gemini-2.5-Pro by 2.0 points and Vision-DeepResearch-30B-A3B by 0.4 points; since 0.4 points is roughly one question here, we treat \method{} as competitive with rather than superior to that system, and regard the 27.0- and 31.0-point margins as the robust gains.

\begin{table}[t]
\centering
\small
\caption{Accuracy (\%) on LiveVQA; best and second-best results are bold and underlined.}
\label{tab:livevqa}
\begin{tabular}{@{}>{\raggedright\arraybackslash}p{0.45\columnwidth}%
                  >{\raggedright\arraybackslash}p{0.27\columnwidth}r@{}}
\toprule
Model & Method & LiveVQA \\
\midrule
GPT-5~\cite{singh2026openaigpt5card} & Agentic + Web search& 73.3 \\
Gemini-2.5-Pro~\cite{comanici2025gemini25pushingfrontier} & Agentic + Web search& 76.0 \\
Gemini-2.5-Flash~\cite{comanici2025gemini25pushingfrontier} & Agentic + Web search& 73.0 \\
Claude-4-Sonnet~\cite{anthropic2025claude4} & Agentic + Web search & 69.7 \\
Claude-3.7-Sonnet~\cite{anthropic2025claude37} & Agentic + Web search& 72.0 \\
Vision-DeepResearch-30B-A3B~\cite{huang2026visiondeepresearchincentivizingdeepresearchcapability} & Agentic + Web search& \underline{77.6} \\
Vision-DeepResearch-8B~\cite{huang2026visiondeepresearchincentivizingdeepresearchcapability} & Agentic + Web search& 76.7 \\
OpenSearch-VL-32B~\cite{opensearchvl} & Agentic + Web search& 70.5 \\
OpenSearch-VL-30B-A3B~\cite{opensearchvl} & Agentic + Web search& 67.4 \\
WebWatcher-32B~\cite{geng2025webwatcherbreakingnewfrontier} & Agentic + Web search& 58.7 \\
MMSearch-R1-7B~\cite{wu-etal-2026-mmsearch} & Agentic + Web search& 48.4 \\
\midrule
Qwen3-VL-8B~\cite{bai2025qwen3vltechnicalreport} & direct, no search & 47.0 \\
Qwen3-VL-8B~\cite{bai2025qwen3vltechnicalreport} & \method{}, no search & 51.0 \\
Qwen3-VL-8B~\cite{bai2025qwen3vltechnicalreport} & \method{} + Web search & \textbf{78.0} \\
\bottomrule
\end{tabular}
\end{table}

\begin{table}[t]
\centering
\setlength{\tabcolsep}{3.5pt}
\caption{Accuracy (\%) on BrowseComp (BC), GAIA, and XBench. Values follow the source-specific protocols described in the text; \method{} variants appear below the separator.}
\label{tab:other-results}
\begin{tabular}{@{}lccc@{}}
\toprule
Method & BC & GAIA & xbench \\
\midrule
OpenAI o4-mini~\cite{openai2025o4mini} & \textbf{28.3} & 55.8 & \textbf{67.0} \\
Claude-4-Sonnet~\cite{anthropic2025claude4} & 12.2 & 57.6 & 64.0 \\
Kimi-K2~\cite{kimiteam2026kimik2openagentic} & 14.1 & \textbf{57.7} & 50.0 \\
DeepSeek-R1~\cite{Guo_2025} & 8.9 & 30.3 & 55.0 \\
Nemotron-3-Nano~\cite{nvidia2025nemotron3nanoopen} & 10.6 & 50.5 & 55.0 \\
ASearcher-QwQ-32B~\cite{gao2025turnsunlockinglonghorizonagentic} & 5.2 & 52.8 & 42.0 \\
WebDancer-QwQ-32B~\cite{wu2025webdancerautonomousinformationseeking} & 3.8 & 51.5 & 39.0 \\
WebSailor-72B~\cite{li2025websailornavigatingsuperhumanreasoning} & 12.0 & 55.4 & 55.0 \\
DeepMiner-32B~\cite{tang2025turnlimitstrainingdeep} & 21.2 & 54.4 & 53.0 \\
Qwen3-8B + ReAct & 4.8 & 29.1 & 25.6 \\
WebExplorer~\cite{liu2025webexplorerexploreevolvetraining} (Qwen3-8B SFT+RL) & 15.7 & 50.0 & 53.7 \\
Qwen3-32B~\cite{yang2025qwen3technicalreport} + ReAct & 1.6 & 29.7 & 33.7 \\
\midrule
\method{} (Qwen3-8B) & 10.5 & 43.7 & 45.0 \\
\method{} (Qwen3-8B RL) & 17.8 & 53.4 & 56.0 \\
\method{} (Qwen3-32B) & 19.5 & 55.0 & 58.0 \\
\bottomrule
\end{tabular}
\end{table}

\subsection{Same-Backbone Ablation}
\label{sec:ablation}
Table~\ref{tab:qwen3-8b} isolates the effect of the control architecture and the RL objective. All three configurations share the same prompt backbone, tool permissions, retrieval backend, and budget of at most 50 LLM turns per episode. The differing search counts (3.7 vs.\ 12.9 vs.\ 13.2) are \emph{emergent behavior} rather than an experimental degree of freedom: the single agent stops searching early, whereas the dual-role loop keeps retrieving until the graph pins down a candidate. The first comparison therefore measures the end-to-end effect of the dual-role architecture, which raises accuracy from 2.7\% to 26.9\%; because search persistence is itself induced by that architecture, this margin reflects role decomposition together with the resulting stopping behavior rather than role decomposition in isolation. The second comparison is tighter: holding the architecture and the mean search count essentially fixed (12.9 vs.\ 13.2), RL raises accuracy from 26.9\% to 35.9\% while reducing mean generated tokens per rollout from 1,878 to 1,689, so the 9.0-point gain is attributable to the RL objective rather than to more retrieval.

\begin{table}[t]
\centering
\caption{Same-backbone ablation on BrowseComp-Plus (830 questions; three-seed mean). Tokens and searches are means per rollout, all variants use a 50-turn budget, and accuracy standard deviations are below 0.5 percentage points.}
\label{tab:qwen3-8b}
\small
\setlength{\tabcolsep}{4pt}
\begin{tabular}{lccc}
\toprule
Control scheme & Acc. & Tok./rollout & Searches \\
\midrule
Single-agent & 2.7 & 3245 & 3.7 \\
Dual-role, no RL & 26.9 & 1878 & 12.9 \\
Dual-role + RL & \textbf{35.9} & \textbf{1689} & 13.2 \\
\bottomrule
\end{tabular}
\end{table}

\subsection{Training Dynamics and Error Analysis}
Figure~\ref{fig:train} shows that reward rises sharply after step 50 and stabilizes at $0.15$--$0.28$ by step 100. Response length falls from about 1,900 to 1,600 tokens before stabilizing near 1,650--1,750, while mean searches increase slightly from 12.9 to 13.1--13.3, indicating that RL reallocates generation toward short queries and compact graph edits rather than longer traces. Two failures dominate: the executor may loop on query reformulations before finding a candidate, and the verifier may fail to open a promising result. The graph exposes these as repeated searches with no new tag and useful retrieval with no accepted edge; rejected spans, unresolved conflicts, and premature answers are likewise observable.

\section{Discussion}
The ablation suggests that the gain comes from changing the retrieval--reasoning interface rather than adding parametric knowledge. A monolithic agent must repeatedly recover the current candidate, verified constraints, and provenance from an expanding trace, whereas \method{} externalizes these variables: the executor sees a compact snapshot, the verifier performs local page inspection, and the policy proposes grounded edits. For $D$ documents, $S$ spans, and $C$ claims, graph storage is $O(D+S+C+|E|)$ and sparse in practice, and the executor snapshot scales as $O(C+|\tags|)$ rather than with the full retrieved text $O(\sum_d|d|)$. Scaling beyond modest constraint sets will require candidate pruning, source deduplication, and hierarchical graph summaries that retain provenance.

\section{Conclusion}
\method{} constructs a typed, span-grounded evidence graph through separate search and evidence-recording roles. At matched budget, role decomposition improves Qwen3-8B from 2.7\% to 26.9\% on BrowseComp-Plus, and joint RL reaches 35.9\% with fewer generated tokens. The same loop reaches 78.0\% on LiveVQA, supporting structured evidence construction as an alternative to unstructured retrieval traces.

\bibliographystyle{ACM-Reference-Format}
\bibliography{references}

\end{document}